\documentclass[12pt,twoside]{article} 
\usepackage[utf8]{inputenc}

\usepackage{amsfonts}
\usepackage{amsmath}
\usepackage{amsthm,amssymb}
\usepackage{cite}
\usepackage{a4}
\usepackage{graphicx}
\usepackage{booktabs}

\newcommand{\I}{\mathrm{i}}        
\newcommand{\E}{\mathrm{e}}        
\newcommand{\D}{\mathrm{d}}        

\newcommand{\sign}{\operatorname{sign}} 

\newcommand{\Det}{\operatorname{det}}

\newcommand{\End}{\operatorname{End}}   
\DeclareMathOperator{\tensor}{\otimes}

\renewcommand{\Im}{\operatorname{Im}} 
\renewcommand{\Re}{\operatorname{Re}} 

\newcommand{\tr}{\operatorname{tr}} 

\newcommand{\ch}{\operatorname{ch}}  
\newcommand{\sh}{\operatorname{sh}}

\newcommand{\As}{\mathcal{A}}
\newcommand{\Bs}{\mathcal{B}}
\newcommand{\Cs}{\mathcal{C}}
\newcommand{\Ds}{\mathcal{D}}

\newcommand{\Us}{\mathcal{U}}

\newcommand{\XXZ}{X\!X\!Z}
\newcommand{\XXX}{X\!X\!X}

\newcommand{\TQ}{\emph{TQ}}

\newcommand{\magnon}{m}

\theoremstyle{plain}
\newtheorem*{example*}{Example}
\newtheorem*{conjecture*}{Conjecture}

\theoremstyle{definition}

\newlength{\HFPP}       \HFPP5.4mm
\makeatletter
\@addtoreset{equation}{section}
\makeatother

\begin{document}

\thispagestyle{empty}

\begin{center}

{\Large {\bf Functional Bethe Ansatz Methods for the Open XXX Chain\\}}

\vspace{7mm}

{\large Holger Frahm, 
Jan H.\ Grelik, 
Alexander Seel and 
Tobias Wirth
}

\vspace{5mm}

Institut f\"ur Theoretische Physik, Leibniz Universit\"at Hannover,\\
Appelstr. 2, 30167 Hannover, Germany\\[2ex]

\vspace{20mm}

{\large {\bf Abstract}}

\end{center}

\begin{quote}
  We study the spectrum of the integrable open $\XXX$ Heisenberg spin chain
  subject to non-diagonal boundary magnetic fields.  The spectral problem for
  this model can be formulated in terms of functional equations obtained by
  separation of variables or, equivalently, from the fusion of transfer
  matrices.  For generic boundary conditions the eigenvalues cannot be
  obtained from the solution of finitely many algebraic Bethe equations.
  Based on careful finite size studies of the analytic properties of the
  underlying hierarchy of transfer matrices we devise two approaches to
  analyze the functional equations.  First we introduce a truncation method
  leading to Bethe type equations determining the energy spectrum of the spin
  chain.  In a second approach the hierarchy of functional equations is mapped
  to an infinite system of non-linear integral equations of TBA type.  The two
  schemes have complementary ranges of applicability and facilitate an
  efficient numerical analysis for a wide range of boundary parameters.  Some
  data are presented on the finite size corrections to the energy of the state
  which evolves into the antiferromagnetic ground state in the limit of
  parallel boundary fields.

\end{quote}

\clearpage

%
\section{Introduction}
%

The solution of the spectral problem for integrable models constructed within
the framework of the Quantum Inverse Scattering method (QISM) is facilitated
by a collection of various very powerful tools, commonly subsumed as Bethe
ansatz methods.  In Bethe's original work on the spin-$1/2$ Heisenberg chain
\cite{Bethe31} eigenfunctions were found by means of an ansatz for solutions
of Schrödinger's equation in terms of scattering states of $\magnon$ magnons
created from the completely polarized state as a reference which led to a
system of coupled algebraic equations for $\magnon$ parameters.  The roots of
these Bethe equations correspond to particular eigenstates of the model in the
$\magnon$-magnon sector.  Within the QISM this approach was put on an
algebraic basis.  For representations of a Yang-Baxter algebra the hamiltonian
is identified as one member of a family of commuting operators generated by
the transfer matrix.  The eigenstates of this transfer matrix are obtained by
the action of creation operators on the ferromagnetic reference state.  The
creation operators are functions of the roots of the Bethe equations.

This need for an eigenstate which is sufficiently simple to guess, however,
limits the use of both the coordinate and the algebraic Bethe ansatz.  For
other cases, e.g.\ for models where the underlying symmetry is realized in a
non-compact way without a highest weight state \cite{Sklyanin85,ByTe06} or
situations where the $U(1)$ symmetry of the bulk is broken by interactions or
boundary terms \cite{Nepomechie02,MuNeSh06a} different approaches are needed:
within the class of functional Bethe ansatz methods, e.g. Baxter's method of
commuting transfer matrices, Separation of Variables (SoV), or the fusion algebra
\cite{Baxter82,Sklyanin92,KlPe92}, relations between elements of the
Yang-Baxter algebra are derived which eventually lead to functional equations
for the eigenvalues of the transfer matrix.  These relations have to be solved
based on the analytic properties of the eigenvalues, e.g.\ distribution of
zeros and poles, and their asymptotic behaviour.  In certain cases this
approach may lead to Bethe equations similar to those found within the
coordinate or algebraic Bethe ansatz.  In some cases where this is not
possible, the functional relations have been brought to a form which allows to
express them in terms of non-linear integral equations.  From the analysis of
these equations it has been possible to gain important insights into the
properties of the ground state and low lying excitations as well as the
thermodynamics for certain models, see e.g.\ \cite{Kluemper04}.

In this paper we employ some of these functional methods to study the spectrum
of the isotropic spin-$1/2$ Heisenberg chain with open boundary conditions where
the symmetry of the bulk is broken due to non-parallel boundary magnetic
fields.  The model is given by the hamiltonian
\begin{equation}\label{hamil}
  \begin{aligned}
      \mathcal{H}_{\XXX} = \sum_{j=1}^{L-1}&\Big[\sigma_j^x\sigma_{j+1}^x + \sigma_j^y\sigma_{j+1}^y +\sigma_j^z\sigma_{j+1}^z \Big] +L\\
      +&\frac{\I}{\alpha^-} \tanh\beta^- \sigma_1^z + \frac{\I}{\alpha^-\ch\beta^-}\big(\ch\theta^-\sigma_1^x + \I\sh\theta^- \sigma_1^y \big)\\
      +&\frac{\I}{\alpha^+} \tanh\beta^+ \sigma_L^z + \frac{\I}{\alpha^+\ch\beta^+}\big(\ch\theta^+\sigma_L^x + \I\sh\theta^+ \sigma_L^y \big)\quad,
    \end{aligned}
\end{equation}
where $\sigma_j^\alpha$, $\alpha=x,y,z$ denote the Pauli matrices acting on
the space of states of a spin-$1/2$ at site $j$ and $\alpha^\pm$, $\beta^\pm$,
$\theta^\pm$ parametrize the boundary fields acting on sites $L$ and $1$
respectively.  

The hamiltonian \eqref{hamil} is obtained from a transfer matrix based on a
representation of Sklyanin's reflection algebra \cite{Sklyanin88} which extends
the QISM to systems with open boundary conditions.
Thereby the integrability of the model is established.  At the same time,
however, the fact that the ferromagnetically polarized state with all spins up
is not an eigenstate of $\mathcal{H}_{\XXX}$ for generic boundary fields
prevents the application of the coordinate or algebraic Bethe ansatz to the
solution of the spectral problem.  In a previous paper \cite{FSW08} we have
used Sklyanin's Separation of Variables method to address this problem and
have derived difference equations, so-called \TQ-equations, which are
satisfied by all eigenvalues of the transfer matrix.  To access properties of
the system in the thermodynamic limit $L\to\infty$, however, the
characterization of the $Q$-functions appearing in these equations was
incomplete.  This is a familiar situation arising in the actual computation of
spectral properties: to extract useful information from the Bethe equations
one always needs some additional insights into the behaviour of their
solutions as the system sizes varies, in particular for those corresponding to
states with low energies or to the equilibrium state at finite temperature.
This requirement appears to limit several other attempts to study spin chains
with non-diagonal boundary terms based on e.g.\ representations of a
$q$-Onsager algebra \cite{BaKo07} or on an alternative use of the Yang-Baxter
algebra due to Galleas \cite{Galleas08}.  Here we tackle this difficulty by
identifying the analytical properties of the objects appearing in the
\TQ-equations and the related fusion hierarchy from finite size studies.

Our paper is organized as follows: in Section \ref{IBC} we present a brief
account of the construction of integrable boundary conditions within the
QISM. We recall previous results obtained from SoV and by fusion of transfer
matrices with higher dimensional auxiliary spaces.  The spectral parameter in
the $\TQ$-equations derived within the SoV approach is restricted to a finite
lattice of points which prevents the use of the functional methods for their
solution.  On the other hand, within the fusion procedure an equivalent
$\TQ$-equation with continuous arguments is obtained assuming that a certain 
limit of the fused transfer matrices exists for infinite dimensional auxiliary 
space (see also \cite{YaNeZh06,NiWiFr09}). Based on this equivalence we perform 
a finite size study of the analytical properties of the transfer matrix eigenvalues
and of the $Q$-functions in Section \ref{QBAE}.  This allows one to derive a finite
system of \lq truncated\rq\ Bethe equations which we solve numerically for 
selected eigenstates of (\ref{hamil}).
In Section \ref{NLIE} we use our finite size data to rewrite the fusion
hierarchy in terms of non-linear integral equations of TBA type whose solution
determines one selected eigenstate of (\ref{hamil}).
The paper ends with a summary of our results and some concluding remarks.


\section{Integrable Boundary Conditions} \label{IBC}
The construction of integrable systems involving boundaries within the QISM was initiated by Sklyanin \cite{Sklyanin88}.
It is valid for a general class of integrable systems characterized by
an $R$-matrix of difference form $R(\lambda,\mu) =$ \mbox{$R(\lambda-\mu)$}$
\in \End(V\tensor V)$ 
which satisfies the Yang-Baxter equation 
\begin{equation} \label{YBE}
R_{12}(\lambda-\mu)\,R_{13}(\lambda-\nu)\,R_{23}(\mu-\nu)
= R_{23}(\mu-\nu)\,R_{13}(\lambda-\nu)\,R_{12}(\lambda-\mu)\quad .
\end{equation}
The indices of $R_{jk}$ denote the embedding where $R$ acts non-trivially on
the tensor product of vector spaces $V_1\tensor V_2\tensor V_3$.  For the
hamiltonian \eqref{hamil} we need the well-known $6$-vertex model solution
\begin{equation} \label{Rmatrix}
R(\lambda) =
  \begin{pmatrix}
     a(\lambda) & 0 & 0 & 0 \\
     0 & b(\lambda) & c(\lambda) & 0 \\
     0 & c(\lambda) & b(\lambda) & 0 \\
     0 & 0 & 0 & a(\lambda)
  \end{pmatrix} \qquad , \qquad
  \begin{aligned}
     a(\lambda) &= \lambda+\I\\
     b(\lambda) &= \lambda\\
     c(\lambda) &= \I
  \end{aligned}
\end{equation}
of the Yang-Baxter equation \eqref{YBE} with $V_j=\mathbb{C}^2$. 
Each solution $R(\lambda)$ fixes the structure
constants of the related Yang-Baxter algebra  
\begin{equation}\label{YBA}
R_{12}(\lambda-\mu) T_1(\lambda) T_2(\mu) = T_2(\mu) T_1(\lambda)R_{12}(\lambda-\mu)
\end{equation}
with generators $T^\alpha_{\phantom{x}\beta}(\lambda)$, $\alpha,\beta=1,2$.
$T_1(\lambda) = T(\lambda) \tensor I$, $T_2(\lambda) = I \tensor T(\lambda)$
are the embeddings of the monodromy matrix $T(\lambda)$ in the product of
auxiliary spaces $V_1\otimes V_2$, they satisfy the inversion formula
\begin{equation}\label{inversion}
T(\lambda)=\begin{pmatrix}A(\lambda)&B(\lambda)\\C(\lambda)&D(\lambda)\end{pmatrix}\quad ,\quad
T^{-1}(\lambda) =\frac{1}{(d_qT)(\lambda-\I/2)}\sigma^y T^t(\lambda-\I)\sigma^y\quad .
\end{equation}
The scalar factor $(d_qT)(\lambda)=A(\lambda+\I/2)D(\lambda-\I/2) -
B(\lambda+\I/2)C(\lambda-\I/2)$ is the central element of the Yang-Baxter
algebra and is known as quantum determinant. The superscript $t$ denotes a transposition
and can be extended to the $j$th auxiliary space  by $t_j$ .

Sklyanin's construction of open spin chains is based on the representations of
two algebras 
$\Us^+(\lambda)$ and $\Us^-(\lambda)$
defined by the relations
\begin{equation}\label{leftalg}
R_{12}(\lambda-\mu)\Us_1^-(\lambda) R_{12}(\lambda+\mu-\I)\Us_2^-(\mu)=
\Us_2^-(\mu)R_{12}(\lambda+\mu-\I)\Us_1^-(\lambda)R_{12}(\lambda-\mu)
\end{equation}
\begin{equation}\label{rightalg}
R_{12}(\mu-\lambda)\Us_1^{+ t_1}(\lambda) R_{12}(-\lambda-\mu-\I)\Us_2^{+ t_2}(\mu)=
\Us_2^{+ t_2}(\mu)R_{12}(-\lambda-\mu-\I)\Us_1^{+ t_1}(\lambda)R_{12}(\mu-\lambda)
\end{equation}
We shall call 
$\Us^+(\lambda)$ and $\Us^-(\lambda)$
right and left reflection algebras
respectively. Considering their product in auxiliary space the trace  
\begin{equation}\label{transfermatrix}
\tau(\lambda) = \tr \Us^{+}(\lambda)\,\Us^{-}(\lambda)
\end{equation}
defines the transfer matrix as the central object under consideration
because it generates with $[\tau(\lambda),\tau(\mu)]=0$ a commuting family of
operators which can be simultaneously diagonalized.  

The explicit construction of integrable open boundary conditions for models
arising from the Yang-Baxter algebra with the $R$-matrix (\ref{Rmatrix})
starts with the $2\times2$ matrix
\begin{equation}\label{K-Nepo}
  K(\lambda,\pm)=\frac{1}{\alpha^\pm \ch\beta^\pm } 
   \begin{pmatrix}
       \alpha^\pm \ch\beta^\pm+\lambda \sh\beta^\pm &   \lambda\,\E^{\theta^\pm} \\
       \lambda\,\E^{-\theta^\pm} &   \alpha^\pm \ch\beta^\pm- \lambda\sh\beta^\pm
  \end{pmatrix}
\end{equation}
independently found by \cite{VeGo93,GhZa94}, considered in the rational limit \cite{Kulish1995} and used here in a parametrization introduced by Nepomechie \cite{Nepomechie04}.
It constitutes the known $c$-number representations $K^{+}(\lambda) =
K(\lambda+\I/2 ,+)$ and $K^{-}(\lambda) = K(\lambda-\I/2,-)$ of the
reflection algebras \eqref{leftalg}, \eqref{rightalg} with the properties $\tr K(\lambda,\pm)=2$ and $K(0,\pm)=I_2$.
Following Sklyanin we choose
\begin{equation} \label{fullrepminus}
\Us^{-}(\lambda) = T(\lambda-\I/2) K(\lambda-\I/2,-) \sigma^y T^t(-\lambda-\I/2)\sigma^y=\begin{pmatrix}
\As^-(\lambda) & \Bs^-(\lambda) \\
\Cs^-(\lambda) & \Ds^-(\lambda)
\end{pmatrix}
\end{equation}
as representation incorporating the inversion formula \eqref{inversion}. This
leads to an explicit representation of the transfer matrix with normalization
condition $\tau(\I/2)= (d_qT)(-\I/2)$ responsible for an additional factor of
$1/2$ in a similar decomposition of the right reflection algebra 
\begin{equation}\label{decompplus}
\Us^{+}(\lambda) = \frac12 K(\lambda+\I/2,+)=
\begin{pmatrix}
\As^+(\lambda) & \Bs^+(\lambda) \\
\Cs^+(\lambda) & \Ds^+(\lambda)
\end{pmatrix}\quad .
\end{equation}
Then the hamiltonian \eqref{hamil} is connected to the transfer matrix by
$\mathcal{H}_{\XXX}=\I\partial\ln\tau(\I/2)$.


\subsection{Quantum Determinants} \label{QD}
Analogously to the quantum determinant of the Yang-Baxter algebra there exists
similar objects for the reflection algebras. 
Quantum determinants play a crucial role when applying functional methods to
solve the spectral problem.  For the left reflection algebra it is defined
according to \cite{Sklyanin88} reading with the projector $P_{12}^-$ onto the
singlet in $V_1\otimes V_2$
\begin{equation}\label{quantendetminus}
(\Delta_q^- \Us)(\lambda) = \tr_{12} P_{12}^- \Us_1^-(\lambda-\I /2)
R_{12}(2\lambda-\I )\Us_2^-(\lambda+\I /2) \quad . 
\end{equation}

To express $(\Delta_q^- \Us)(\lambda)$ in terms of the generators
$\As^-(\lambda)$, $\Bs^-(\lambda)$, $\Cs^-(\lambda)$ and $\Ds^-(\lambda)$ it is
instructive to use the combinations 
\begin{equation}
\widetilde{\Ds}^-(\lambda) \equiv 2\lambda \Ds^-(\lambda) -
\I \,\As^-(\lambda) \,, \quad  
\widetilde{\Cs}^-(\lambda) \equiv (2\lambda+\I ) \Cs^-(\lambda)
\end{equation}
borrowed from the algebraic Bethe ansatz. Then the suggestive form of the quantum
determinant reads  
\begin{equation}
(\Delta_q^- \Us)(\lambda) = \As^-(\lambda+\I /2)\,\widetilde{\Ds}^-(\lambda-\I /2) - \Bs^-(\lambda+\I /2)\,\widetilde{\Cs}^-(\lambda-\I /2) \quad .
\end{equation}

Thus in case of the $c$-number representation $K(\lambda-\I /2,-)$ connected to the left reflection algebra $\Us^-(\lambda)$ the relation
\begin{equation}
(\Delta_q^- K)(\lambda-\I /2,-) = 2(\lambda-\I ) \Det K(\lambda,-) = -2(\lambda-\I )\frac{(\lambda-\alpha^-) (\lambda+\alpha^-) }{ (\alpha^-)^2 }
\end{equation}
holds. Note that this connection is only valid for the shifted argument $\lambda-\I /2$ because the arising expressions in \eqref{quantendetminus} are no longer of difference form.
As the quantum determinant respects co-multiplication, applying it to the full
representation \eqref{fullrepminus} of the left reflection algebra with the monodromy $T(\lambda)$ yields
\begin{equation} \label{qdetref}
(\Delta_q^- \Us)(\lambda) = (d_q T)(\lambda-\I /2) \, (\Delta_q^- K)(\lambda-\I /2,-) \,
(d_q T)(-\lambda-\I /2) \quad .
\end{equation}

The right reflection algebra can be treated in a similar way. We may leave with the suggestive form of the result
\begin{equation}
(\Delta_q^+ \Us)(\lambda) = {\Ds}^+(\lambda-\I /2)\,\widetilde{\As}^+(\lambda+\I /2) - \Bs^+(\lambda-\I /2)\,\widetilde{\Cs}^+(\lambda+\I /2) \quad  .
\end{equation}
Again, we used some suitable combinations reading
\begin{equation}
\widetilde{\As}^+(\lambda) \equiv -2\lambda\As^+(\lambda) -\I\Ds^+(\lambda) \quad ,\quad
\widetilde{\Cs}^+(\lambda) \equiv (-2\lambda+\I ) \Cs^+(\lambda)
\end{equation}
where in case of the $c$-number representation $K(\lambda+\I /2,+)$ to the algebra $\Us^+(\lambda)$ the quantum determinant takes the form
\begin{equation}
(\Delta_q^+ K)(\lambda+\I /2,+) = -2(\lambda+\I ) \Det K(\lambda,+) = 2(\lambda+\I )\frac{(\lambda-\alpha^+) (\lambda+\alpha^+) }{ {(\alpha^+)}^2 }\quad .
\end{equation}


\subsection{Separation of Variables} 
For diagonal boundary matrices the spectrum of the model has be obtained with
the algebraic Bethe ansatz by action of the operators $\Bs^-(\lambda)$ on the
completely polarized pseudo vacuum $|0\rangle$ \cite{Sklyanin88}.  In the case
of generic boundary conditions, $|0\rangle$ is not an eigenstate of the
transfer matrix and this approach is not possible.
Instead one can follow Sklyanin's functional approach \cite{Sklyanin92} based
on the operator valued zeros $\widehat{x}_j$ of $\Bs^-(\lambda)$,
$j=1,\ldots,L$.  Studying representations of the reflection algebra on a space
of symmetric functions of the eigenvalues $x_j$ of these operators one obtains
the \TQ-equation for the eigenvalues $\Lambda$ of the transfer matrix
\cite{FSW08},
\begin{equation}
\label{TQsep}
 \Lambda(x_j)\, Q(x_j) =\frac{(-1)^L}{2x_j}\Delta^+(x_j)
                    Q(x_j+\I)
   + \frac{(-1)^L}{2x_j}\Delta^-(x_j)
                   Q(x_j-\I)\quad .
\end{equation}
The coefficients $\Delta^\pm(\lambda)$ factorize the quantum determinant to the transfer matrix \eqref{transfermatrix} according to $(\Delta_q^+\Us)(\lambda)\,(\Delta_q^-\Us)(\lambda)=-\Delta^+(\lambda-\I/2)\Delta^-(\lambda+\I/2)$ . Generalizing the transfer matrix to include inhomogeneous shifts
parametrized by $L$ lattice parameters $s_j$ the eigenvalues of
$\widehat{x}_j$ are found to be $x_j^\pm=s_j\pm\I/2$ and the functions
$\Delta^\pm$ are explicitly given as
\begin{equation}
\begin{aligned}
\Delta^-(\lambda) &= (\lambda+\I/2)\frac{\left(\lambda+\alpha^+-{\I}/{2}\right)
                   \left(\lambda+\alpha^--{\I}/{2}\right)}{\alpha^+\alpha^-}\prod_{\ell=1}^L(\lambda-s_\ell+\I/2)(\lambda+s_\ell+\I/2)\quad,\\
\Delta^+(\lambda) &= (\lambda-\I/2)\frac{\left(\lambda-\alpha^++{\I}/{2}\right)
                   \left(\lambda-\alpha^-+{\I}/{2}\right)}{\alpha^+\alpha^-}\prod_{\ell=1}^L(\lambda-s_\ell-\I/2)(\lambda+s_\ell-\I/2)\quad.
\end{aligned}
\end{equation}
Note that only the diagonal parameters $\alpha^\pm$ of the boundary matrices
enter in these equations.  To obtain the spectrum for non-diagonal boundary
fields corresponding to values of the parameters $\beta^\pm$ and $\theta^\pm$
they have to be complemented with information on the asymptotic behaviour of
$\Lambda(\lambda)$ at large $|\lambda|\gg1$:  from the construction of the transfer 
matrix one easily obtains \cite{FSW08}
\begin{equation}\label{AsymptoticOfEigenvalue}
  \Lambda(\lambda) \sim \frac{(-1)^L\ch\phi}{\alpha^+\alpha^-}\,\lambda^{2L+2}\,,
  \quad
  \ch\phi \equiv \frac{\sh\beta^+\sh\beta^-+\ch(\theta^+-\theta^-)}{
                    \ch\beta^+\ch\beta^-}\quad .
\end{equation}
Hence the parameter $\phi$ is sufficient to characterize the influence of the
non-diagonal boundary fields.  As a change in the sign of one
$\alpha$-parameter can be absorbed into the corresponding $\beta\to-\beta$ and
$\theta\to\theta\pm\I\pi$ with the mapping $\ch\phi\to-\ch\phi$ the complete
parameter range of \eqref{hamil} is governed by $\ch\phi>0$ along with the
cases $\alpha^+/\I,\alpha^-/\I>0$ and $\alpha^-/\I < 0 <\alpha^+/\I$.  
As a simultaneous change $\alpha^\pm \to-\alpha^\pm$ formally reverses all spatial 
directions we do not need to consider the range $\alpha^+/\I,\alpha^-/\I<0$.

The $x_j^\pm$ are singular points of the difference equation (\ref{TQsep}),
i.e.\ simple roots of the coefficients $\Delta^\pm(x_j^\pm)=0$.  Therefore,
the $Q$-functions can be eliminated from (\ref{TQsep}) in favour of a
functional equation for $\Lambda(x)$ valid on the discrete set
$g=\{x_j^\pm\}$. It has been shown \cite{FSW08} that this equation yields the
complete spectrum of the transfer matrix for small lattices.  In the
homogeneous limit $s_j\to s$ it becomes
\begin{equation}\label{griddeterminant}
\Lambda_g(s+\I/2)\Lambda_g(s-\I/2) = \frac{(s-\alpha^+)(s-\alpha^-)}{\alpha^+\alpha^-}\frac{(s+\alpha^+)(s+\alpha^-)}{\alpha^+\alpha^-}
\frac{(s^2+1)^{2L+1}}{4s^2+1}\quad .
\end{equation}
Here he subscript $g$ emphasizes that (\ref{griddeterminant}) has been
derived for the eigenvalue $\Lambda(\lambda)$ with arguments $\lambda$ taken from the set
$g$ and therefore holds only up to terms which vanish as \mbox{$(\lambda-s)^{2L+2}$}.
Still, neglecting this fact and taking $s$ to be a continuous variable one can
solve for $\partial\ln\Lambda_g(s)$ by Fourier transformation \cite{FSW08}. A particular state can be selected by imposing constraints on the analytical properties of $\Lambda_g$. For example the ground state eigenvalue has no zeros in the strip $|\Im z|<1/2$.
Evaluating the result at the point $\I/2$ (c.f.\ \eqref{energyvalue}) this
leads to the correct bulk and boundary contribution $E_g$ to the ground state
energy ($\psi(x)$ is the digamma function)
\begin{equation}\label{Lambdaongrid}
  \begin{split}
E_g\equiv\I\partial\ln{\Lambda_g}({\I}/{2}) =
  & 
  +\psi({|\alpha^+|}/{2})- \psi\big(({|\alpha^+|}+{1})/{2}\big)+{1}/{|\alpha^+|}
  \\[.3em]
  & 
  +\psi({|\alpha^-|}/{2})- \psi\big(({|\alpha^-|}+{1})/{2}\big)+{1}/{|\alpha^-|}
  \\[.3em]
&+    \pi -2 \ln 2-1 +(2-4\ln2)L\quad.
   \end{split}
\end{equation}
Finite size corrections of order $\mathcal{O}(1/L)$ which would capture
correlations between the two ends of the chain are beyond this approach.

\subsection{Fusion Procedure}
The so-called fusion procedure grants the possibility to easily obtain
$R$-matrices and boundary matrices of higher dimensions obeying a Yang-Baxter
or reflection equation respectively.  
In case of the $R$-matrix this procedure is applicable to the auxiliary, the
quantum space, and even both. 
Furthermore the associated transfer matrices are not independent from each
other but satisfy functional relations called {fusion hierarchies}. 

An $R$-matrix of dimension $k/2$ in auxiliary space and a spin-$1/2$ representation in quantum space is given by e.g.\ \cite{Nepomechie02} reading
\begin{equation}\label{fusionRmatrix}
  R_{\langle1\cdots k\rangle \ell} (\lambda) = P^+_{1\cdots k} R_{1\ell}(\lambda) R_{2\ell} (\lambda+\I ) \cdots R_{k\ell}(\lambda+(k-1)\I ) P^+_{1 \cdots k} \quad.
\end{equation}
Here, the projector $P^+$ is defined as
\begin{equation}
P^{+}_{1\cdots n} = \frac{1}{n!} \sum_{\sigma\in\mathfrak{S}_n} P_{\sigma} 
\end{equation}
where the sum runs over all permutations $\sigma\in \mathfrak{S}_n$ of the
symmetric group 
and $P_\sigma$ is the permutation operator reordering the positions in the
space $(\mathbb{C}^2)^{\tensor n}$ according to $\sigma$. 
The fused boundary matrices are defined in \cite{Zhou96b} following Mezincescu
and Nepomechie \cite{MeNe92a} who carried out the first fusion step  
\begin{equation}
K_{\langle12\rangle}^+(\lambda) = P_{12}^+ K_2^+(\lambda+\I/2)R_{12}(-2\lambda-2\I ) K_1^+(\lambda-\I/2 ) P_{12}^+
\end{equation}
similar to \eqref{fusionRmatrix} explicitly. Utilizing the co-multiplication property one also finds
\begin{equation}
\Us^{-}_{\langle12\rangle}(\lambda) = P^+_{12} \,\Us^{-}_1(\lambda) R_{12}(2\lambda)\,\Us^{-}_2(\lambda+\I )P^+_{12}
\end{equation}
yielding a transfer matrix with a spin-$1$ auxiliary space
$
\tau_{\langle12\rangle}(\lambda) \equiv \tr_{\langle12\rangle} K^+_{\langle12\rangle}(\lambda)\, \Us^{-}_{\langle12\rangle}(\lambda)
$.
With these definitions Mezincescu and Nepomechie \cite{MeNe92a} showed the
fusion formula for the transfer matrix of the open boundary model 
\begin{equation}\label{fusion_obc_tau}
\tau_{\langle12\rangle}(\lambda-\I/2)=-(2\lambda-\I )(2\lambda+\I )\,\tau(\lambda-\I/2 )\tau(\lambda+\I/2)-\tfrac{1}{4}(\Delta_q^+K)(\lambda+{\I }/{2},+)\,(\Delta_q^-\Us)(\lambda)
\end{equation}
depending on the quantum determinants of both reflection algebras and the
original transfer matrix $\tau(\lambda)$ from \eqref{transfermatrix}.  This
result can be rewritten by absorbing the scalar prefactor in front of the
$\tau$'s into the definition of the fused transfer matrix giving
\begin{equation}\label{fusion_lvl1}
t_2(\lambda-\I )=t_1(\lambda-\I )t_1(\lambda)-\delta(\lambda)\quad ,\quad \tau(\lambda+\I/2)=\frac{(-1)^L}{\alpha^+\alpha^-}t_1(\lambda)
\end{equation}
with the scalar function $\delta(\lambda)$ on the RHS reading
\begin{equation}\label{delta}
\delta(\lambda) = \frac{(\lambda^2+1)^{2L+1}}{4\lambda^2+1} (\lambda-\alpha^+)(\lambda+\alpha^+)(\lambda-\alpha^-)(\lambda+\alpha^-)
\quad .
\end{equation}

Extending this procedure to higher dimensional auxiliary
spaces\cite{Zhou96b,Nepomechie04} the arising transfer matrices $t_k$ for
integer $k$ are finally related to each other through the fusion hierarchy
\begin{equation}\label{fusion-hierarchy}
t_k\big(\lambda-\I(k-1) \big)=t_{k-1}\big(\lambda-\I(k-1) \big)t_1(\lambda)-\delta(\lambda)t_{k-2}\big(\lambda-\I(k-1) \big)\;,\;  k=2,3,\ldots
\end{equation}
with $\delta(\lambda)$ given in \eqref{delta} and $t_0(\lambda) \equiv 1$.
Note that we do not have to distinguish between the transfer matrices
$t_k(\lambda)$ and their eigenvalues because we are dealing with commuting
quantities sharing a common system of eigenfunctions. Thus in the following we
will use the notation $t_k$ also for the eigenvalues.
Again, the fusion hierarchy needs to be completed with the asymptotic
behaviour $t_k(\lambda) \sim a_k \lambda^{(2L+2)k}$ of the eigenvalues of the
fused transfer matrices: solving a recursion relation following from
\eqref{fusion-hierarchy} and the asymptotic \eqref{AsymptoticOfEigenvalue},
\eqref{fusion_lvl1} of the eigenvalue $\Lambda(\lambda)$ of $\tau(\lambda)$ as
a starting value one obtains
\begin{equation}\label{asymptotix}
t_k(\lambda) \sim a_k \lambda^{(2L+2)k} \quad ,\quad
a_k = \frac{1}{2^k}\frac{\sh\big((k+1)\phi\big)}{\sh \phi}\quad.
\end{equation}


\subsection{Equivalence of TQ-Equations}\label{fe}
The fusion hierarchy \eqref{fusion-hierarchy} can formally be solved for the shifted eigenvalue $t_1(\lambda-\I/2)$ related to spin-$1/2$ reading
\begin{equation}
t_1(\lambda-\I/2)= \frac{t_k(\lambda+\I/2-\I (k+1)+\I)}{t_{k-1}(\lambda+\I/2-\I k)} + \delta(\lambda-\I/2)\frac{t_{k-2}(\lambda+\I/2-\I (k-1)-\I)}{t_{k-1}(\lambda+\I/2-\I k)}
\end{equation}
matching the general form of a $TQ$-equation \cite{YaNeZh06}.  Indeed,
renormalization of the fused transfer matrices $t_k(\lambda)$ according to
\begin{equation}\label{tauk}
  t_k(\lambda) = a_k \Big[\prod_{\ell=1}^{k-1} (\lambda+\I\ell)^{2L+1}\,
           \prod_{\ell=1}^{k-1} (\lambda+\I\ell +{\I}/{2} )^{-1}\Big]\, \tau_k(\lambda)
\end{equation}
and assuming the limit $\lim_{k\to\infty} \tau_k(\lambda-\I k-{\I}/{2})\equiv
q(\lambda)$ to exist yields a difference equation
\begin{equation}
\label{qbae:TQ1}
\begin{aligned}
\Lambda(\lambda) \,q(\lambda) =& \frac{(-1)^L}{\alpha^+\alpha^-}\frac{\E^\phi}{2\lambda}\left(\lambda-{\I}/{2}\right)^{2L+1} q(\lambda+\I)\\
                 &+ \frac{(-1)^L}{\alpha^+\alpha^-}\frac{\E^{-\phi}}{2\lambda}
\left(\lambda+{\I}/{2}\right)^{2L+1}
                   \Big[\prod_{\sigma=\pm}(\lambda-{\I}/{2}-\alpha^\sigma)(\lambda-{\I}/{2}+\alpha^\sigma)\Big]
q(\lambda-\I)
\end{aligned}
\end{equation}
which fixes the eigenvalue $\Lambda(\lambda)$ of $\tau(\lambda)$.  With the
transformation
\begin{equation}
\label{qbae:qQ}
q(\lambda) =  (\mp1)^{\I\lambda} \E^{\I\lambda\phi}Q(\lambda) \left[ \Gamma(-\I \alpha^++{1}/{2}+\I\lambda) \Gamma(-\I \alpha^-+{1}/{2}+\I\lambda)
  \right]^{-1}\,
\end{equation}
we can absorb the exponential dependence on $\phi$ into $Q(\lambda)$
recovering the $TQ$-equation already obtained from the functional Bethe ansatz
\cite{FSW08} reading 
\begin{equation}
\label{qbae:TQ0}
\begin{aligned}
\pm\Lambda(\lambda)\, Q(\lambda) =& \frac{(-1)^L}{2\lambda\alpha^+\alpha^-}\left(\lambda-{\I}/{2}\right)^{2L+1}
                   \left(\lambda-\alpha^++{\I}/{2}\right)
                   \left(\lambda-\alpha^-+{\I}/{2}\right) Q(\lambda+\I)\\
   &+ \frac{(-1)^L}{2\lambda\alpha^+\alpha^-}\left(\lambda+{\I}/{2}\right)^{2L+1}
                   \left(\lambda+\alpha^+-{\I}/{2}\right)
                   \left(\lambda+\alpha^--{\I}/{2}\right)Q(\lambda-\I)\quad.
\end{aligned}
\end{equation}
In addition to the explicit appearance of $\alpha^\pm$ the boundary conditions
enter this equation through the large-$\lambda$ behaviour
\eqref{AsymptoticOfEigenvalue} of the eigenvalue $\Lambda(\lambda)$ and
$\tau(\lambda)$ respectively.  

For boundary parameters giving
$\ch\phi=\pm1$ equation~(\ref{qbae:TQ0}) can be solved by an even polynomial
$Q(\lambda)=\prod_{j=1}^M (\lambda-\lambda_j)(\lambda+\lambda_j)$ where the
$\lambda_j$ are roots determined from the Bethe equations
\begin{equation}
\label{qbae:bae}
  \left(\frac{\lambda_j+{\I}/{2}}{\lambda_j-{\I}/{2}}\right)^{2L+1}
      \frac{\lambda_j+\alpha^+-{\I}/{2}}{\lambda_j-\alpha^++{\I}/{2}}\,
      \frac{\lambda_j+\alpha^--{\I}/{2}}{\lambda_j-\alpha^-+{\I}/{2}}
  = -\prod_{k=1}^M\frac{\lambda_j-\lambda_k+\I}{\lambda_j-\lambda_k-\I}\,
                 \frac{\lambda_j+\lambda_k+\I}{\lambda_j+\lambda_k-\I}\quad .
\end{equation}
This case corresponds to boundary fields which can be dealt with by means of
the algebraic Bethe ansatz.  On the other hand, for $\ch\phi\ne\pm1$ no simple
ansatz for $Q(\lambda)$ is known. However, for large $|\lambda|$ it has to grow
exponentially $Q \sim \exp(-\I\phi \lambda)$ according to \eqref{qbae:qQ}. 


\section{Truncated Bethe Equations} 
\label{QBAE}
After explicitly carrying out the large-$k$ limit of the auxiliary space
dimension for the transfer matrix $t_k$ we are able to derive truncated Bethe 
equations in the non-diagonal boundary parameter range of $\phi \in
\mathbb{R}$. Studying the root distributions of $t_k$ for small system sizes
yields information about the general root structure and it is possible to
approximately treat the system with a finite number of zeros. With this
knowledge on the analytical properties of the $q$-functions \eqref{qbae:TQ1} leads to
truncated Bethe equations. As a remark in the case of the $\XXZ$ model with
diagonal boundaries and a special choice of boundary parameters the fusion
hierarchy truncates exactly and the problem was already solved in
\cite{Zhou95} and \cite{Nepomechie02} for non-diagonal boundaries
respectively.
\begin{figure}[t]
  \begin{center}
     \includegraphics[height=9cm]{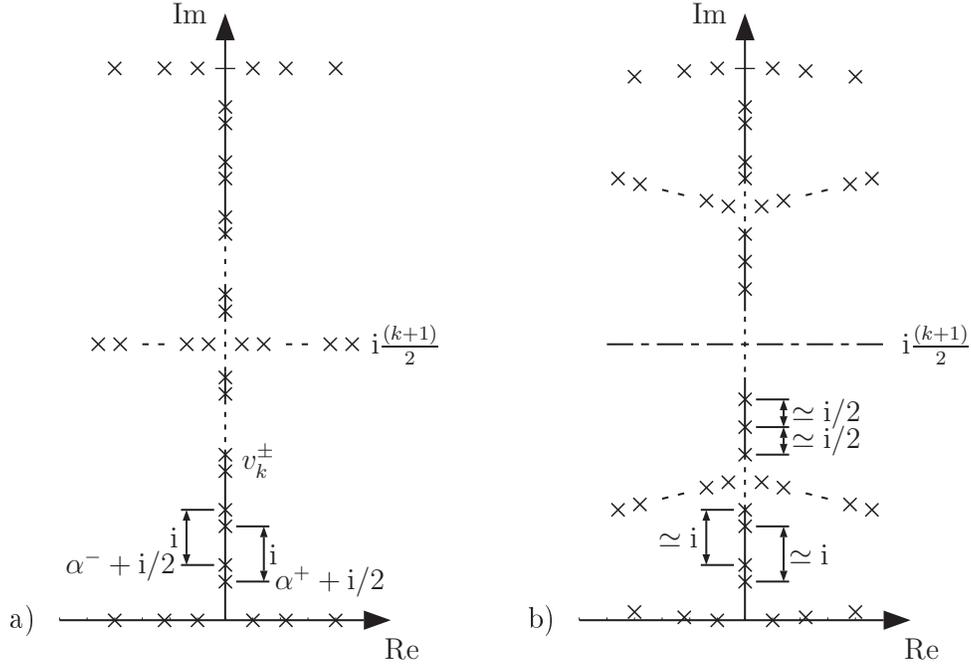}
  \end{center}
\caption{Typical root distributions for the $A$-state of 
$\tau_k(\lambda-\I k -{\I}/{2})$
for an even system size $L$ and $\Im\alpha^-,\Im\alpha^+>1/2$ (a) in the Bethe ansatz solvable case of $\ch\phi = 1$  and (b) in the non-diagonal case of $\ch\phi>1$. Both distributions are symmetric to $\lambda=\I(k+1)/2$ .}\label{qbae:fig}
\end{figure}


\subsection{Iteration of the Fusion Equations}\label{qbae:Iteration}
For systems with a few lattice sites ($L<14$) the polynomial eigenvalues of
the transfer matrix can be calculated explicitly either by exact formulas or
numerical determination of the coefficients.  Then the fusion hierarchy
\eqref{fusion-hierarchy} can be used to compute the corresponding
$\tau_k(\lambda)$ up to a specific fusion level $k$.

The roots of these functions $\tau_k(\lambda)$ are found\footnote{ This may
  not be true for all states. In cases where the \lq Bethe ansatz-part\rq\ of
  the roots contains one or more string solutions the roots of $q(\lambda)$
  may extend into the half plane $\Re\lambda<0$.  } 
to be in the strip $-k-1/2\le \Im\lambda \le 1/2$ and are symmetric with
respect to the line $\Im\lambda=-k/2$.
As a consequence the zeros of $\tau_k(\lambda-\I k -\I/2)$ in \eqref{tauk},
which are relevant in the limit $k\to\infty$ for the $q$-function, emerge in
the region $0\le \Im\lambda < (k+1)/2$ as denoted in Figure~\ref{qbae:fig} for
the Bethe ansatz solvable case (a) and for non-diagonal boundaries (b). Here
we will concentrate on two distinguished states which we label by $A$ and $B$:
in the diagonal limit $\ch\phi=\pm1$ amenable to the algebraic Bethe ansatz
the first state $A$ turns into the singlet ground state of the antiferromagnetic 
chain whereas the $B$-state describes in this limit the fully
magnetized state.  Below we will study these states for boundary parameters
$\Im\alpha^\pm>1/2$ which excludes boundary bound states.

For the Bethe ansatz solvable case in Figure~\ref{qbae:fig}a we observe in the
$A$-state a distribution of zeros on the real axis, which are the known
$2\times L/2$ Bethe roots for this sector.  In addition the $q$-function has
roots on the imaginary axis which form a half-infinite lattice of spacing $\I$
starting at the points $\lambda=\alpha^\pm+{\I}/{2}$.  In the limit
$k\to\infty$ this lattice becomes exact which allows to rewrite
(\ref{qbae:TQ1}) as (\ref{qbae:TQ0}) with polynomial $Q(\lambda)$ as discussed
above.
The extra roots appearing on the symmetry line $\Im\lambda=(k+1)/2$ for finite
$k$ can be neglected in the limit $k\to\infty$. 

In cases 
with no Bethe ansatz the roots on the symmetry line move to branches in the
complex plane which strongly depend on the asymptotic $\ch\phi$.
A typical root configuration for the $A$-state is shown in
Figure~\ref{qbae:fig}b: we find that the number of roots forming these
branches is fixed and their positions in the complex plane show only a slow
variation with respect to the system size.  The additional roots near the real
axis evolve into the $2\times L/2$ solutions of the Bethe equations
(\ref{qbae:bae}) parametrizing the antiferromagnetic ground state in the
limit $\ch\phi\to1$.  The zeros $v_k^\pm$ on the imaginary axis start from
the points $\alpha^\pm+{\I}/{2}$ with exponential accuracy (in $L$) and are
approximately spaced by $\I$ as in the Bethe ansatz solvable case.
However, at some imaginary point depending on the boundary parameters there is
a crossover of the sequence $v_k^\pm$ into a lattice of roots spaced by
$\I/2$.  This lattice is approached with exponential accuracy which suggests
to replace the sequence $v_k^\pm$ beyond some (half) integer position $\I M$ by a
compensating factor accounting for the asymptotics.  This allows to treat the
finitely many remaining roots with imaginary part $<M$ as in the treatise of
the Bethe ansatz solvable case from above.

For the $B$-state a strong size dependence of the branches of extra roots is
found which requires a more careful finite size analysis (see below).


\subsection{Truncation}
The analyticity of the transfer matrix eigenvalues (\ref{qbae:TQ1}) implies
that the roots $q(\lambda_j)=0$ have to satisfy an infinite hierarchy of Bethe
equations
\begin{equation}
\label{qbae:bax}
  \E^{-2\phi}\left(\frac{\lambda_j+{\I}/{2}}{\lambda_j-{\I}/{2}}\right)^{2L+1}
      \prod_{\sigma=\pm}(\lambda_j-{\I}/{2}-\alpha^\sigma)(\lambda_j-{\I}/{2}+\alpha^\sigma)
    = - 
\frac{q(\lambda_j+\I)}{q(\lambda_j-\I)}
\end{equation}
for all $j\in\mathbb{N}$.  In addition the restriction
\begin{equation}
\E^{-2\phi}(\I/2-\alpha^+)(\I/2+\alpha^+)(\I/2-\alpha^-)(\I/2+\alpha^-)=\frac{q(\I)}{q(-\I)}
\end{equation}
emerges from the residue at $\lambda=0$ to vanish.  Our findings on the
asymptotic positions of the roots with $\mathrm{Im}\lambda_j\gg1$ (c.f.\
Figure~\ref{qbae:fig}b), i.e.\ by consecutive integers and half integers on
the imaginary axis suggest us to explicitly deal only with the finite number $N$
of zeros with $\mathrm{Im}\lambda_j< M$ for some sufficiently large $M$
depending on the boundary parameters $\alpha^\pm$ and $\phi$.
Within this approach  the roots with $\Im\lambda_j>M$ cancel (almost) perfectly in \eqref{qbae:TQ1}, i.e.\ we can replace
\begin{equation}
\begin{aligned}
  \frac{q(\lambda+\I)}{q(\lambda)} &\to 
 f(\lambda)
   \prod_{k=1}^N(\lambda-\lambda_k+\I)
   \prod_{\ell=1}^{N-2} \frac{1}{\lambda-\lambda_\ell} \quad ,\\
\frac{q(\lambda-\I)}{q(\lambda)} &\to
\frac{1}{f(\lambda-\I)}
   \prod_{\ell=1}^{N-2}(\lambda-\lambda_\ell-\I)
   \prod_{k=1}^{N} \frac{1}{\lambda-\lambda_k}\quad .
\end{aligned}
\end{equation}
The asymptotic behaviour \eqref{AsymptoticOfEigenvalue} of the eigenvalue $\Lambda(\lambda)$ in the representation \eqref{qbae:TQ1} requires $f(\lambda)\equiv1$. As a result we obtain
\begin{equation}
\label{qbae:TQ1sugg}
\begin{split}
 \Lambda(\lambda)  =& \frac{(-1)^L}{\alpha^+\alpha^-}\frac{\E^\phi}{2\lambda}
\left(\lambda-{\I}/{2}\right)^{2L+1} (\lambda-\lambda_N+\I)(\lambda-\lambda_{N-1}+\I)\prod_{k=1}^{N-2}\frac{\lambda-\lambda_k+\I}{\lambda-\lambda_k}\\
&+ \frac{(-1)^L}{\alpha^+\alpha^-}\frac{\E^{-\phi}}{2\lambda}\left(\lambda+{\I}/{2}\right)^{2L+1}
\frac{\prod_{\sigma=\pm}(\lambda-{\I}/{2}-\alpha^\sigma)(\lambda-{\I}/{2}+\alpha^\sigma)}{(\lambda-\lambda_N)(\lambda-\lambda_{N-1})}
            \prod_{k=1}^{N-2}\frac{\lambda-\lambda_k-\I}{\lambda-\lambda_k}    
\end{split}
\end{equation}
from which we can derive truncated Bethe equations by the requirement of
vanishing residues.  The $N$ remaining roots of $q(\lambda)$ have to satisfy
the system
\begin{multline}
\label{qbae:bax2}
  \E^{-2\phi}\left(\frac{\lambda_j+{\I}/{2}}{\lambda_j-{\I}/{2}}\right)^{2L+1}=\\
      =\frac{(\lambda_j-\lambda_{N-1})(\lambda_j-\lambda_{N-1}+\I)
     (\lambda_j-\lambda_{N})(\lambda_j-\lambda_{N}+\I)}{(\lambda_j-{\I}/{2}-\alpha^+)(\lambda_j-{\I}/{2}+\alpha^+) (\lambda_j-{\I}/{2}-\alpha^-)(\lambda_j-{\I}/{2}+\alpha^-)}
     \prod_{\substack{k=1\\k\not=j}}^{N-2}\frac{\lambda_j-\lambda_k+\I}{\lambda_j-\lambda_k-\I}
\end{multline}
for $j=1,\ldots,N-2$.  For $j=N-1,N$ the equations are ill-defined since the
RHS vanishes identically.  Note, however, that the analyticity of
$\Lambda(\lambda)$ requires the residues of (\ref{qbae:TQ1sugg}) at
$\lambda=\lambda_N$ and $\lambda=\lambda_{N-1}$ to vanish.  This condition can
be met by setting $\lambda_{N-1}\equiv\I M+\I/2$ and $\lambda_{N}\equiv\I
M+\I$ for the (half) integer $M$ chosen for the truncation above as the zeros
$\lambda_{N-2}\approx\I M$ and $\lambda_{N-3}\approx\I M -\I/2$ are located
close but to not precisely on the asymptotic positions.
As a test from the singularity of the eigenvalue $\Lambda$ at $\lambda=0$ the
roots have to obey the restriction 
\begin{equation}
\E^{-2\phi}=\frac
{\lambda_N(\lambda_N-\I)\lambda_{N-1}(\lambda_{N-1}-\I)}
{(\I/2-\alpha^+)(\I/2+\alpha^+)(\I/2-\alpha^-)(\I/2+\alpha^-)}
\prod_{k=1}^{N-2}\frac{\lambda_k-\I}{\lambda_k+\I}\quad .
\end{equation}

Equations~(\ref{qbae:bax2}) can be solved numerically by Newton's algorithm.  For the
$A$-state this is possible due to the fact that the number and position of the
roots forming the branch (c.f.\ Figure~\ref{qbae:fig}b) vary only slowly
with respect to the system size for fixed boundaries.  For the $2\times L/2$
zeros near the real axis one can use the root distribution from the Bethe
ansatz solvable case \eqref{qbae:bae} as starting values.  Doing so one obtains solutions for
up to several thousand lattice sites which can be used to examine e.g.\ the
finite size effects (c.f.\ Figure~\ref{finite-size} in Section~\ref{NLIE}).

\begin{figure}
  \begin{center}
     \includegraphics[height=7cm]{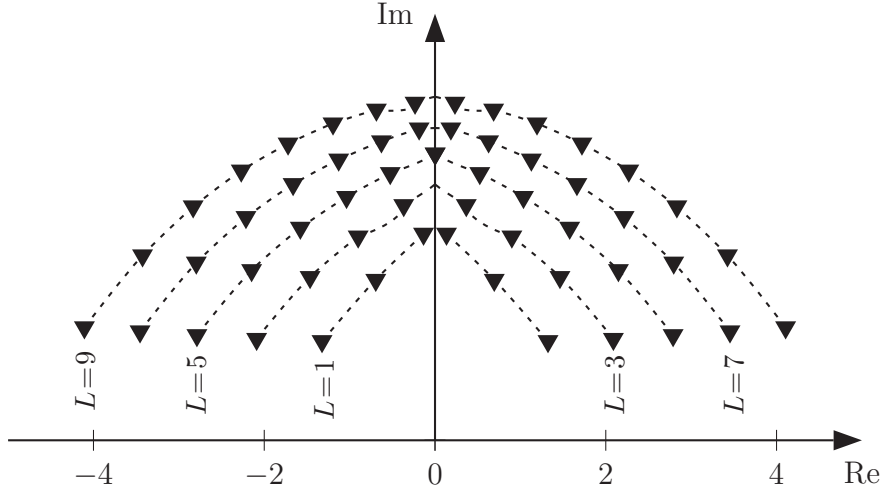}
  \end{center}
\caption{Root distributions for the $B$-state at $\alpha^\pm=11\I/3$ and $\ch\phi=11$. Zeros belonging to fixed lattice sizes $L=1$ to 
$L=9$ 
are each connected by a dashed line as a guidance for the eyes. The roots on the imaginary axis forming an asymptotic lattice of half integers are not displayed.}\label{qbae:ferrostate}
\end{figure}

In the $B$-state the shape and position of the branches as well as the number
of roots forming them strongly varies with the system size which prevents the
derivation of truncated Bethe equations used above.  However, a careful
analysis of the arising root distributions leads to characteristic patterns as
depicted in Figure~\ref{qbae:ferrostate} where we show typical results for
$\alpha$-parameters of equal sign.

\subsection{Scope of Application}
Clearly the method can also be applied to boundary bound states
$0<\Im\alpha^\pm<1/2$ with roots sticking to the points $\I/2-\alpha^\pm$ now
being part of the parametrization \eqref{qbae:TQ1sugg} of the eigenvalue. For
a plain overview of the terminology and parameter range see e.g.\
\cite{SeWi09}.

Note that the derivation of the equations \eqref{qbae:bax2} does not depend on
the signs of the boundary parameters $\alpha^\pm$: 
for boundary fields with opposite signs, $\sign(\Im\alpha^+)=
-\sign(\Im\alpha^-)$, the root configuration corresponding to the $A$-state
differs from the case above by the absence of the additional branches in the
complex plane.  For the $B$-state, branches as in Figure~\ref{qbae:ferrostate} are still present. Therefore, in both cases an analysis of the truncated
Bethe equations is possible along the lines described before.


\section{Y-System and Non-Linear Integral Equations}\label{NLIE}
In order to directly capture the corrections to the bulk and boundary part
\eqref{Lambdaongrid} of the energy eigenvalue of the open spin chain one can
use the methods of complex calculus to utilize the equivalent representation
\cite{KlPe92}
\begin{equation}\label{tsystem}
t_k(\lambda+{\I }/{2})t_k(\lambda-{\I}/{2})= t_{k-1}(\lambda+{\I}/{2}) t_{k+1}(\lambda-{\I}/{2}) 
+ \prod_{\ell=1}^k \delta(\lambda-{\I}/{2}+\I\ell) 
\end{equation}
of the fusion hierarchy \eqref{fusion-hierarchy}. Following the standard
scheme \cite{KuNaSu94} one introduces the combination
\begin{equation} 
y_2(\lambda)=\frac{t_2(\lambda-\I)}{\delta(\lambda)}
\end{equation}
being part of an infinite series $\{y_k\}$ related by functional relations. Once
$y_2$ is known the eigenvalue of the underlying integrable model can be
calculated from the lowest level
\begin{equation}\label{funceqLambda}
  \Lambda(\lambda+\I/2)\,\Lambda(\lambda-\I/2) = \frac{\delta(\lambda)}{(\alpha^+\alpha^-)^2} \big(1+ y_2(\lambda)\big)
\end{equation}
of the fusion hierarchy \eqref{tsystem} e.g.\ by Fourier techniques. Note the
invariance of the functional equation \eqref{funceqLambda} with respect to
$\Lambda\to-\Lambda$ or $\alpha^\pm\to-\alpha^\pm$.


\subsection{Y-System and Fourier Transformation}
The aforementioned infinite system of functional relations is denoted as
$Y\!$-system and can be derived for products of the transfer matrix
eigenvalues. It is sometimes called the universal form of the
TBA-equations\cite{KuNaSu94} and follows directly from the fusion hierarchy
\eqref{tsystem}. By defining
\begin{equation}
y_k(\lambda)\equiv\frac{t_{k-2}(\lambda-\I(k-2)/2) t_k(\lambda-\I k/2)}{\prod_{\ell=1}^{k-1}\delta(\lambda-\I k /2+\I\ell)}
\end{equation}
and explicitly calculating $(1+y_{k+1})(1+y_{k-1})$ by making use of the fusion relation \eqref{tsystem} one finds for integers $k$
\begin{equation}\label{Ysystem}
y_k(\lambda-\I/2)\,y_k(\lambda+\I/2) = \big(1+y_{k-1}(\lambda)\big)\big(1+y_{k+1}(\lambda)\big) 
\end{equation}
with $y_1\equiv0$.  Note that a simultaneous scaling of $t_k(\lambda)$ and
$\delta(\lambda)$ from \eqref{fusion-hierarchy}
\begin{equation}
  \label{Yscaleinvariance}
  \begin{gathered}
  t_1(\lambda)\rightarrow \frac{t_1(\lambda)}{p(\lambda)} 
  \;\;, \quad 
  \delta(\lambda) \rightarrow \frac{\delta(\lambda)}{p(\lambda)p(\lambda-\I)}
  \;\;, \quad 
  t_k\big(\lambda-\I(k-1)\big)\rightarrow \frac{t_k\big(\lambda-\I(k-1)\big)}{\prod_{\ell=0}^{k-1}p(\lambda-\I\ell)}
\end{gathered}
\end{equation} 
with any function $p(\lambda)$ leaves the $Y$-system (\ref{Ysystem})
invariant.
In view of this fact we choose $p(\lambda)=(2\lambda+\I )^{-1}$ which allows
to consider polynomial $\delta(\lambda)$ and $t_k(\lambda)$ rather than
rational functions.  Thus the zeros and poles of the $y$-functions can be
identified with the zeros of $t_k(\lambda)$ and $\delta(\lambda)$,
respectively.  
Taking the logarithmic derivative of the $Y$-system (\ref{Ysystem}) and
Fourier transforming
\begin{equation}\label{Fouriertransform}
  \widehat{f}(k) = \int\displaylimits_{-\infty}^{\infty}\!\!\D x \,\E^{-\I k
    x} f(x)\quad ,\quad f(x) =
  \int\displaylimits_{-\infty}^{\infty}\!\!\frac{\D k}{2\pi} \, \E^{\I k x}
  \widehat{f}(k) 
\end{equation}
the imaginary shifts occurring on the LHS can be removed leading to an infinite
set of non-linear integral equations (NLIE) for the $y$-functions evaluated on
the real line \cite{Takahashi71a},
\begin{equation}
\begin{aligned}
  \ln y_2(x) &= d_2(x) + s * \ln(1+y_3)\quad, \\
  \ln y_k(x) &= d_k(x) + s * \ln(1+y_{k-1}) + s * \ln(1+y_{k+1}) \quad,
  \quad k>2\quad.
\end{aligned}
\end{equation}
Here $(s*f)$ denotes a convolution of $f$ with the kernel
\begin{equation}
  s(x)= \frac{1}{2\ch(\pi x)}\quad,\quad \widehat{s}(k) =
  \frac{1}{2\ch(k/2)}\quad. 
\end{equation}
As a consequence of the zeros and poles of the functions $y_k$ their
logarithmic derivatives are not analytic which produces the additional
contributions $d_j(x)$ due to the residue theorem. The determination of
these model-dependent driving terms is the challenging component of the
problem.  To reduce these terms we consider the toy equation
\begin{equation}\label{modeleq}
y(x+\I/2)y(x-\I/2)=F(x)
\end{equation}
for $y(x)$ where $F(x)$ is given explicitly with constant asymptotic and the
auxiliary condition $y(\pm\chi)=0$ for some $\chi\in
\{z\in\mathbb{C}\big|0\leq\Im z<1/2\}$.  Taking the logarithmic derivative
one is led to 
\begin{equation}
\int\displaylimits_{-\infty}^\infty\!\!\!\D x\,\E^{-\I k x}\partial\ln y(x\pm\I/2)=
\E^{\mp k/2}\Big[\widehat{\partial\ln y}(k)\mp2\pi\I\,\E^{\mp\I k \chi}\Big]\quad .
\end{equation}
Switching to Fourier space we can solve \eqref{modeleq} for
$\widehat{\partial\ln y}(k)$ using the integral representation of the digamma
function yielding
\begin{equation}\label{elementary}
\int\displaylimits_{-\infty}^{\infty}\!\!\frac{\D k}{2\pi}   \frac{\sh(\nu
  k/2)}{\ch(k/2)}\E^{\I k x}= 
-\frac{1}{2\pi\I}\partial_x\ln\bigg[ \frac{\ch(\pi x)-\sin(\pi \nu/2)}{\ch(\pi
  x)+\sin(\pi \nu/2)} \bigg]\quad . 
\end{equation}
Identifying $\nu=1-2\chi/\I$, inverse Fourier transforming and integration
with respect to the variable $x$ yields
\begin{equation} \label{drivingterms}
\ln y(x) = C + \ln\bigg[ \frac{\ch(\pi x)-\cos(\I\pi \chi)}{\ch(\pi x)+\cos(\I\pi \chi)} \bigg]
+(s*\ln F)(x)
\end{equation}
where the integration constant $C$ can be recovered from the asymptotic
condition of the initial equation \eqref{modeleq}.\\

Turning to the actual zeros of the $y$-functions we find that each $y_{k>2}$
has a double zero at $\lambda=0$ whereas $y_2$ exhibits a $(2L+2)$-fold zero
at $\lambda=0$. To proceed we have to choose a particular state for which the
transfer matrix $t_1$ and $y_k$ do not contain further zeros in the strip $|\Im z|
<{1}/{2}$ usually called hole-type solutions.  
Such a state exists for even lattice sites and
opposite signs\footnote{Note that for $\alpha^\pm$-parameters of equal sign 
and small $k$ the branches in Figure~\ref{qbae:fig}b appear as
unwanted hole-type solutions in the considered strip. The same holds in the parameter range $\phi \in \I\mathbb{R}$ for any combination of the $\alpha$-parameters.} 
of the boundary
parameters $\alpha^\pm$ differing in their absolute values only by some finite amount (see below). This state is 
restricted to $\phi\in\mathbb{R}$, i.e.\ $\ch\phi>1$, and in
the diagonal limit of $\phi \rightarrow 0$ it becomes the state with lowest
energy in the sector of vanishing magnetization labeled $A$-state before. 
The corresponding driving terms \eqref{drivingterms} read for each double zero at $x=0$ with $\chi\to0$
\begin{equation}
\ln\bigg[ \frac{\ch(\pi x)-\cos(\I\pi\chi)}{\ch(\pi x)+\cos(\I\pi\chi)} \bigg] \overset{\chi\to0}{\longrightarrow}
\ln\tanh^2|\pi x/2|\equiv\ln b^0_1(x)\quad .
\end{equation}

\begin{figure}[t]
  \begin{center}
    \includegraphics[height=12.5cm]{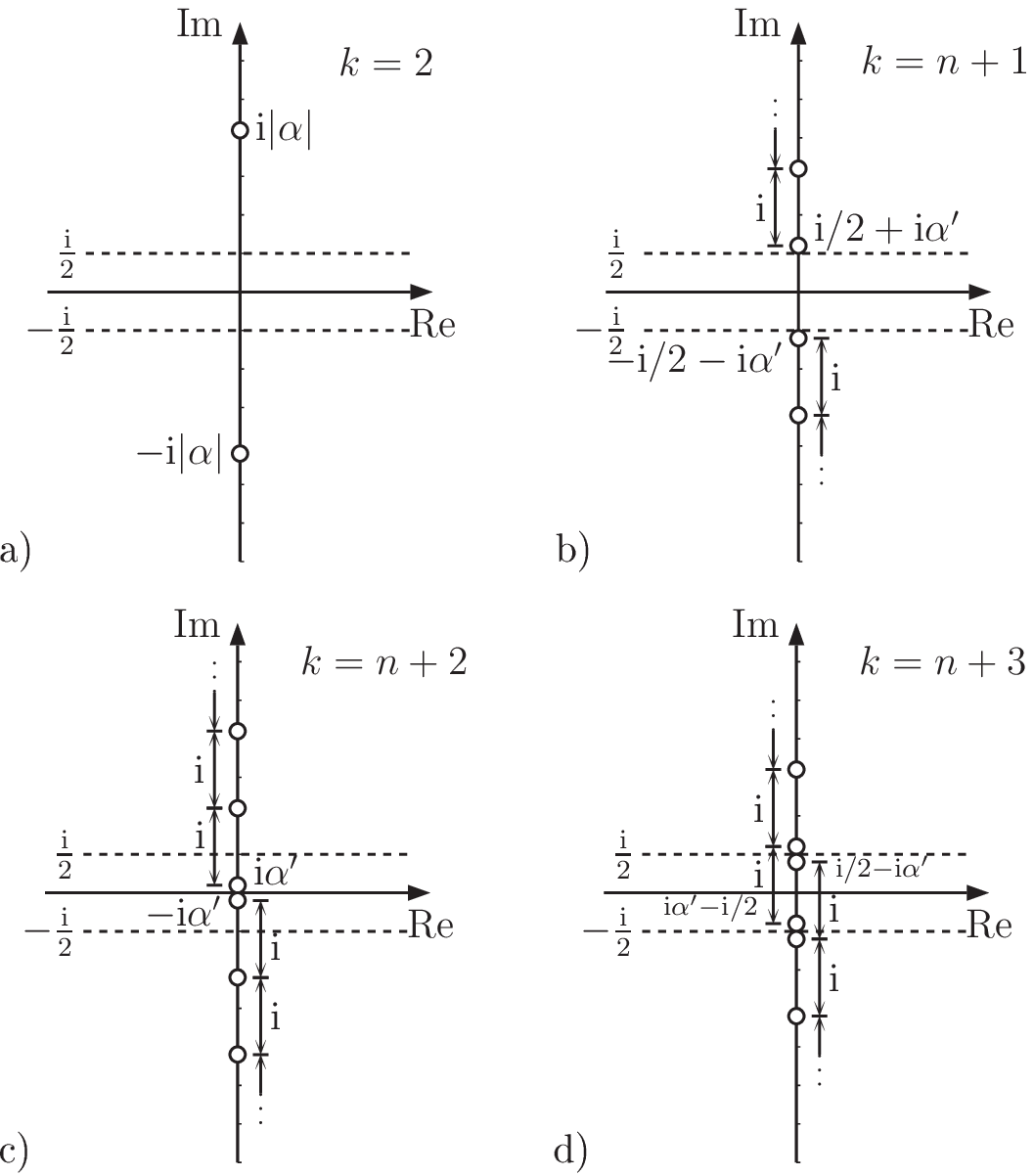}
  \end{center}
  \caption{Exemplary position of poles ($\circ$) of $y_k$ in the complex plane arising from one boundary field $\alpha$ for a) $k=2$, b) $k=n+1$, c) $k=n+2$ and d) $k=n+3$.
  With these examples all different cases of singularities inside the strip $|\Im\lambda|<{1}/{2}$ are already considered. }\label{fig:poles}
\end{figure}

The poles of the $y$-functions can be treated in a similar way. Again, using \eqref{Yscaleinvariance} to scale the transfer matrix properly only the polynomial denominators, e.g.\ $\delta(\lambda+\I/2)\delta(\lambda-\I/2)$  for $y_3$ are responsible for the pole structure with respect to the boundary fields $\alpha^\pm$. The multiple poles at $\lambda=0$ are compensated by the transfer matrices.
Already from the form of the arguments of each $\delta$ given in \eqref{delta} it is clear that the positions of poles vary by $\I/2$ for successive $k$ and two poles of the same $k$ differ by $\I$.
The general scheme is depicted in Figure~\ref{fig:poles}.
Thus using the parametrization $\alpha = \I({n}/{2}+\alpha^\prime)$ with $n\in\mathbb{N}_0$ fixed and $0<\alpha^\prime<{1}/{2}$ the function $y_k(\lambda)$ shows poles at $\lambda=\pm \I \alpha^\prime$  for indices $k\in \{n+2\mathbb{N}\}$ and poles at $\lambda=\pm(\I/2-\I\alpha^\prime)$ for indices $k\in \{n+1+2\mathbb{N}\}$. The corresponding driving terms arise from \eqref{elementary} by inverting the sign on the RHS accounting for poles rather than zeros yielding negative logarithms of
\begin{align}
  h_\text{c}(x) & \equiv
  \frac{\ch(\pi x)-\cos(\pi \alpha^\prime)}{\ch(\pi x)+\cos(\pi \alpha^\prime)}
  && \text{for }  k\in\{n+2\mathbb{N}\} \\
   h_\text{s}(x) & \equiv
   \frac{\ch(\pi x)-\sin(\pi \alpha^\prime)}{\ch(\pi x)+\sin(\pi \alpha^\prime)} 
   && \text{for } k\in\{n+1+2\mathbb{N}\} \quad .
\end{align}
Note that only the modulus of the parameter $\alpha$ enters and the terms $h_\text{c}$ and $h_\text{s}$ alternate for successive $k$. In summary a single zero at $\lambda=0$ gives a driving term of $\ln \tanh|{\pi x}/{2}|$ and due to logarithmic derivative the multiplicity is only reflected in the integer prefactor.
Rearranging all driving terms in matrix form
\begin{equation}
  b_k^n(x) = \left(\begin{array}{c|ccccccc}
       &              1               &       2        &       3       &      4        & \multicolumn{2}{c}{\cdots}  & k \\[.5ex]
\hline\\[-1.7ex]
    0  & \tanh^{2}|{\pi x}/{2}| & h_\text{c}(x)  & h_\text{s}(x) & h_\text{c}(x) & \multicolumn{2}{c}{\cdots}  &\\ 
    1  &              1               &       1        & h_\text{c}(x) & h_\text{s}(x) & \multicolumn{2}{c}{\cdots}  &\\ 
    2  &              1               &       1        &       1       & h_\text{c}(x) & \multicolumn{2}{c}{\cdots}  &\\ 
\vdots &              \vdots          &       \vdots   &      \vdots   & \vdots        & \multicolumn{2}{c}{\ddots}  &\\ 
     n &                        &        &       &        & & \\ 
   \end{array}   \right)  
\end{equation}
allows us to write the non-linear integral equations in a compact form.  For
boundary parameters $\alpha^\pm=\pm\I(n^\pm/2+\alpha_\pm^\prime)$ with
$n^+=n^-=n \in \mathbb{N}_0$ we find for an asymptotic condition $\ch\phi \geq 1$
of the transfer matrix the infinite set of equations
\begin{equation}\label{NLIEobc}
  \begin{aligned}
    \ln {y_2(x)} &=   (L+1)\ln b_1^0(x) - \ln b_2^{n^+}(x) - \ln b_2^{n^-}(x) + s * \ln(1+y_3) \\[1ex]
    \ln y_k(x) &= \ln b_1^0(x) - \ln b_k^{n^+}(x) - \ln b_k^{n^-}(x) \\[.5ex]
     &\quad +  s * \ln(1+y_{k-1}) + s * \ln(1+y_{k+1}) \quad , \quad k>2 
  \end{aligned}
\end{equation}
for an even $L$.  With the asymptotic condition $b^n_k(x)\to1$ for
$x\to\pm\infty$ no further integration constants have to be considered as the
constant asymptotic
\begin{equation}\label{asymptoticsol}
y_k\sim y_k^\infty=\frac{\sh\big((k-1)\phi\big)\sh\big((k+1)\phi\big)}{\sh^2\phi}
\end{equation}
already satisfies the hierarchy \eqref{NLIEobc} in the limit of $x\rightarrow
\pm\infty$.  

To solve an infinite system of equations such as (\ref{NLIEobc}) numerically a
controlled scheme for its truncation is needed.  In the standard TBA approach
\cite{Takahashi99}, e.g.\ for periodic boundary conditions, such an
approximation can be based on the fact that (i) only the first few of the NLIE
contain a non-zero driving term and (ii) the asymptotic solution
(\ref{asymptoticsol}) solves the NLIE without driving terms.  Replacing
$y_k(x)$ by this constant asymptotic for some $k$ chosen sufficiently large
one is left with a finite set of NLIE which can be solved numerically very
efficiently.

In the case of open boundary conditions the NLIE contain driving terms for all
levels $k$ and therefore the asymptotics (\ref{asymptoticsol}) determine only
the large-$x$ behaviour.  Nevertheless, our numerical results using the
constant asymptotic of $y^\infty_k$
as an approximative limiting function seems to assure convergence of the
system reasonably well.  In the special cases of diagonal boundaries $\phi=0$
we find that it is better to scale the constant asymptotic by the necessary
driving terms in order to gain higher accuracy or use less equations without
losing accuracy.

For asymptotic parameters $\phi\neq 0$ the constant solution
\eqref{asymptoticsol} grows exponentially with $k$ and numerical limitations
are quickly reached restricting the method to $\ch\phi=\mathcal{O}(1)$.  The
number of non-linear integral equations, necessary for decent accuracy,
produces function values for which more sophisticated numerical treatment is
needed.
In principle, this simple approach which just uses the asymptotic as a limit
function can be improved by considering solutions of \eqref{NLIEobc} in the
limit of $\max\{n^+,n^-\}\ll k$.  It turns out, however, that this more
sophisticated treatment of the asymptotics shown in Appendix \ref{appA} does 
not allow to reduce the number of NLIE substantially and that the restriction 
to $\ch\phi=\mathcal{O}(1)$ prevails.

\subsection{Energy Eigenvalue}
\begin{figure}[t]
\begin{center}
\includegraphics[height=7.6cm]{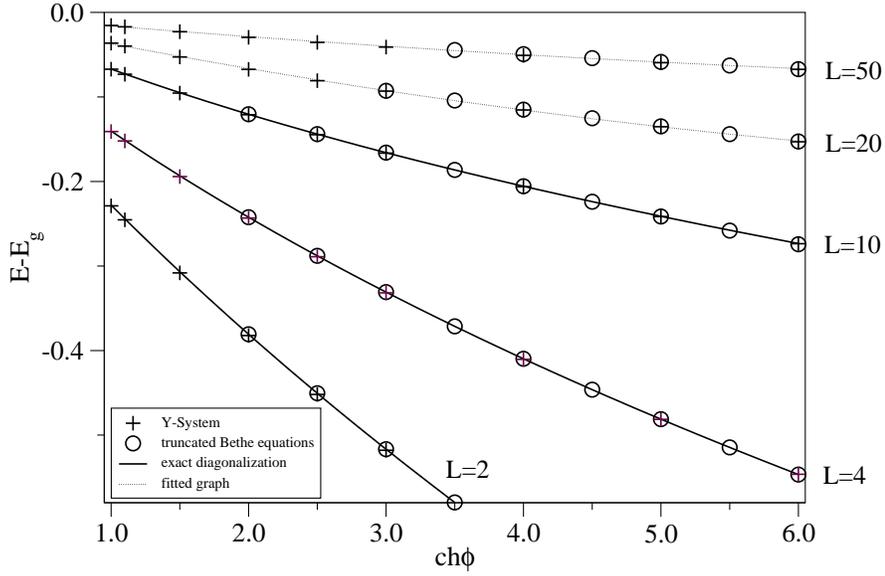}
\end{center}
\caption{Influence of non-diagonal boundaries on the finite size effects to the energy eigenvalue for the $A$-state with parameters $\alpha^+=5\I/3$ and $\alpha^-=-7\I/5$ of different sign and lattice sizes $L=2,4,10,20,50$. For large asymptotics $\ch\phi$ the calculations are performed with the truncated Bethe equations ($\circ$) matching for intermediate values of $\ch\phi$ the results from the $Y\!$-system ($+$). The latter one deals with low asymptotics down to $\ch\phi=1.$ Solid lines are data from exact diagonalization for small system sizes $L=2,4,10$.}
\label{finite-size}
\end{figure}

Once the full hierarchy for $y_k(x)$ is solved the function $y_2(x)$ (and its analytical continuation) can be used to determine the polynomial eigenvalue $\Lambda(\lambda)$ of the transfer matrix. Identifying the quantum determinant $\delta(\lambda)=(\alpha^+\alpha^-)^2\Lambda_g(\lambda+\I/2)\Lambda_g(\lambda-\I/2)$ with \eqref{griddeterminant} we see from the lowest level 
\begin{equation} \label{lowestlevelfusion}
  \Lambda(\lambda+\I/2)\,\Lambda(\lambda-\I/2) =  \Lambda_g(\lambda+\I/2)\Lambda_g(\lambda-\I/2)\big(1+ y_2(\lambda)\big)
\end{equation}
of the fusion hierarchy that the corrections to bulk and boundary contributions \cite{FSW08} are contained in $y_2$. The logarithmic derivative of the eigenvalue reads after simple manipulations in Fourier space
\begin{align}\label{eigenvaluefourier}
\partial\ln\Lambda(x) &= \partial\ln\Lambda_g(x) + \int\displaylimits_{-\infty}^\infty\!\!\frac{\D k}{2\pi}\frac{\E^{\I k x}}{2\ch(k/2)}\widehat{\partial\ln(1+y_2)}(k)\\\label{eigenvaluerealspace}
&=\partial\ln\Lambda_g(x) -{\pi}\!\!\int\displaylimits_{-\infty}^\infty\!\!\frac{\D y}{2}\frac{ \sh(\pi(x-y))}{\ch^2(\pi(x-y))}\ln(1+y_2)(y) \quad .
\end{align}
The expression can be used for asymptotics $y_k^\infty>1$ (i.e.\ $\ch\phi\geq1$) and opposite signs of the boundary fields $\alpha^\pm = \pm\I({n^\pm}/{2}+\alpha^{\pm\prime})$ with $0<\alpha^{\pm\prime}<{1}/{2}$ and $n^\pm\in\mathbb{N}_0$ fixed.\footnote{
In our derivation of the non-linear integral equations \eqref{NLIEobc} we have 
restricted the boundary fields by choosing $n^+=n^-$.  In numerical solutions 
of the equations we find, however, that the error due to the violation of this 
constraint is of similar order as the one arising from the truncation of the 
infinite hierarchy, see Figure~\ref{finite-size}.} 
However, a change of the sign of one $\alpha$-parameter transforms $\phi\to\phi\pm\I\pi$ but leaves the asymptotics $y_k^\infty$ from \eqref{asymptoticsol} unchanged. This again reflects the invariance of the functional equation \eqref{lowestlevelfusion} with respect to $\Lambda\to-\Lambda$ and $\alpha^\pm\to-\alpha^\pm$.
As a remark 
the polynomial eigenvalue $\Lambda(\lambda)$ can be evaluated explicitly in the limit of $\phi\to\infty$ reading
\begin{equation} \label{exacteigenvalue}
\Lambda(\lambda)= \frac{(-1)^L\ch\phi}{\alpha^+\alpha^-}\big(\lambda^2+1/4\big)^{L+1}\quad .
\end{equation}
Note that the prefactor is recovered from the asymptotic behaviour \eqref{AsymptoticOfEigenvalue} and does not contain the combined limit of $\phi,|\alpha^\pm|\to\infty$ but $\ch\phi/(\alpha^+\alpha^-)$ fixed.
Continuing \eqref{eigenvaluefourier} and \eqref{eigenvaluerealspace} respectively to $x=\I/2$ leaves us with the energy eigenvalue of the hamiltonian \eqref{hamil}
\begin{equation}\label{energyvalue}
E= \I\partial\ln\Lambda(\I/2) = \I\partial\ln{\Lambda_g}({\I}/{2})
-{\pi}\!\!\int\displaylimits_{-\infty}^\infty\!\!\frac{\D y}{2}\frac{ \ch(\pi y)}{\sh^2(\pi y)}\ln(1+y_2)(y)
\quad .
\end{equation}
The bulk and boundary part $E_g\equiv\I\partial\ln{\Lambda_g}({\I}/{2})$ was already specified in \eqref{Lambdaongrid} and remains valid for any boundary conditions and all $0<|\alpha^\pm|<\infty$  as the poles emerging from \mbox{$0<|\alpha^\pm|<1/2$} in both $\Lambda(\lambda)$ and $\Lambda_g(\lambda)$ cancel out. Corrections due to correlations between the boundaries in the finite system are covered by the second term yielding the energy dependence in Figure~\ref{finite-size}. 
Clearly in the limit $\phi\to\infty$ the expression \eqref{exacteigenvalue} diverges when the energy is calculated and reveals all hole-type solutions to accumulate at $\lambda=\pm\I/2$. 

\section{Summary and Outlook}
In this paper we have presented two approaches for the analysis of the
functional equation describing the spectrum of the $\XXX$ spin chain with
non-diagonal open boundary conditions (\ref{hamil}).  Based on the integrable
structures underlying this model the \TQ-equation arise both from Sklyanin's
separation of variables and from the fusion procedure for transfer matrices.
Usually, the solution of the \TQ-equation for lattice models with compact
realization of the symmetry can be re-expressed in terms of algebraic Bethe
equations which are obtained in a straightforward way starting with a
polynomial ansatz for the $Q$-functions.  For generic boundary conditions such
an ansatz does not satisfy the requirements on the asymptotic behaviour of the
eigenvalues in the present case (see \cite{FSW08,AFOW10}).
Starting from a finite size study of the analytic behaviour of the transfer
matrix eigenvalues we have been able to express the functional equation in a
way which opens the possibility of a numerical treatment for systems sizes
beyond what is accessible to exact diagonalization.  Both the truncated Bethe
equations derived in Section \ref{QBAE} and the non-linear integral equations
presented in Section \ref{NLIE} can be applied to compute the spectrum for
non-diagonal boundary fields parametrized by $\ch\phi>1$.  They are
particularly useful to compute the energy of the state which evolves into the
antiferromagnetic ground state of the chain for diagonal boundary conditions
(the $A$-state).

The truncated Bethe equations are found to work especially well for
$\ch\phi\gg1$ and can deal with arbitrary boundary parameters $\alpha^\pm$.
Going to large system sizes the numerical analysis of the equations is limited
in principle by the necessity to find appropriate starting values for the
Bethe roots.  Still, selected energies can be computed for systems of several
thousand sites.
In the second method of non-linear integral equations the system size enters
merely as a parameter.  At the same time, however, boundary parameters
$\alpha^\pm$ have to be of different sign to meet certain analyticity
requirements.  In addition, the presence of zeros together with the
exponential growth of the asymptotic value of the $y$-functions with the level
$k$ leads to numerical instabilities limiting the use of these equations to
boundary fields with $\ch\phi=\mathcal{O}(1)$.  
Taken together the two methods are complementary allowing to cover
the entire range of boundary parameters with $\ch\phi>1$.  Our numerical data
show 
these approaches agree for intermediate
values of $\ch\phi$.  Further support for their validity comes from comparison
with exact diagonalization for lattices with up to $14$ sites.

At the same time there remain several open problems which we shall address in
the future:
for the $\XXX$ chain (\ref{hamil}) considered in this paper the description of
the root distribution for the state evolving into the fully polarized
ferromagnetic state in the limit of diagonal boundary conditions ($B$-state,
see Figure \ref{qbae:ferrostate}) needs to be modified to extend the non-linear
integral equations approach to this state.
The limitation of the non-linear integral equations for the $A$-state to
boundary parameters satisfying $\ch\phi\gtrsim1$ appears to be a technical
problem which could be resolved eventually by finding an exact truncation of
the infinite hierarchy of equations similar to that for the $sl(2)$ model
\cite{Suzuki98}.
Furthermore, the case $0<\ch\phi<1$ corresponding to Hermitian boundary terms
in (\ref{hamil}) is not covered by our present approaches.  An extension to
this range of parameters may be possible following the treatment of excited
states for the periodic $\XXZ$ chain to deal with hole-type solutions in the
complex strip to be considered for the $Y\!$-System \cite{KRMSS00}.
Similarly, the present restriction to opposite signs of the
$\alpha$-parameters in the non-linear integral equation approach could be
resolved.
Finally, the methods introduced here should be extended to the $\XXZ$ chain
with non-diagonal boundaries.  The corresponding \TQ-equations have already
been derived from the fusion procedure \cite{YaNeZh06}.  Of particular
interest in this model is the expansion of the energy eigenvalues around the
Bethe ansatz solvable case, $\ch\phi=1$, which would allow for a
non-perturbative study of current fluctuations in certain models for diffusion
in one dimension \cite{DeDoRo04,GiEs06}.\\

{\bf Acknowledgements.}  
The authors would like to thank F.H.L.~Essler, A.M.~Grabinski and A.~Kl\"umper
for helpful discussions.
This work has been supported by the Deutsche Forschungsgemeinschaft under
grant numbers FR~737/6 and SE~1742/1-2.  JHG acknowledges support from the NTH
School for Contacts in Nanosystems.

\begin{appendix}
\section{Asymptotic Truncation}\label{appA}
As the system of equations \eqref{NLIEobc} has
alternating driving terms we assume two limiting functions $g_1(x)$ and
$g_2(x)$ to truncate the system by
\begin{equation}
    y_k(x)= y_k^\infty \frac{\tanh^2|{\pi x}/{2}|\, g_1(x)}{b_k^{n^+}(x)\,b_k^{n^-}(x)} \quad ,\quad
    y_{k^\prime}(x)= y_{k^\prime}^\infty \frac{\tanh^2|{\pi x}/{2}|\, g_2(x)}{b_{k^\prime}^{n^+}(x)\,b_{k^\prime}^{n^-}(x)} 
\end{equation}
for successive $k$ and $k^\prime = k+1\gg\max\{n^+,n^-\}$ supported by
numerical observations. Inserting these into \eqref{NLIEobc} we obtain two
coupled non-linear integral equations
\enlargethispage{1\baselineskip}
\begin{equation}
  \begin{aligned}
    \ln g_1(x)&= s *\ln \left( 
    \frac{1+ y_{k-1}^\infty \frac{b^0_1\, g_2}{b_{k-1}^{n^+}b_{k-1}^{n^-}}}{1+y_{k-1}^\infty} \cdot 
    \frac{1+ y_{k+1}^\infty \frac{b^0_1\, g_2}{b_{k+1}^{n^+}b_{k+1}^{n^-}}}{1+y_{k+1}^\infty}   
    \right) \\[1em]
    \ln g_2(x)&= s *\ln \left( 
    \frac{1+ y_{k^\prime-1}^\infty \frac{b^0_1\,g_1}{b_{k^\prime-1}^{n^+}b_{k^\prime-1}^{n^-}}}{1+y_{k^\prime-1}^\infty} \cdot 
    \frac{1+ y_{k^\prime+1}^\infty \frac{b^0_1\,g_1}{b_{k^\prime+1}^{n^+}b_{k^\prime+1}^{n^-}}}{1+y_{k^\prime+1}^\infty}   
    \right)\quad .
  \end{aligned}
\end{equation}
In the limit of $k\rightarrow \infty$ the coupled system linearizes due to the exponentially fast growing asymptotics $y_k^\infty$ into
\begin{equation}
  \begin{aligned}
    \ln g_1(x)  &= 2 s *\ln g_2 + 2s*\ln b_1^0 -2s* \ln b_{k-1}^{n^+}       -2s* \ln b_{k-1}^{n^-} \\ 
    \ln g_2(x)  &= 2 s *\ln g_1 + 2s*\ln b_1^0 -2s* \ln b_{k^\prime-1}^{n^+}-2s* \ln b_{k^\prime-1}^{n^-} \\ 
  \end{aligned}
\end{equation}
and is solvable in Fourier space after differentiating to apply the residue
theorem. The explicit expression reads
\begin{gather}
    g_1(x) = \Bigg[ 2\ch(\pi x)\frac{\ch({\pi x}/{2})}{d_k^{n^+}(x)}\frac{\ch({\pi x}/{2})}{d_k^{n^-}(x)} \Bigg]^2 \quad , \quad
    g_2(x) = \Bigg[ 2\ch(\pi x)\frac{\ch({\pi x}/{2})}{d_{k+1}^{n^+}(x)}\frac{\ch({\pi x}/{2})}{d_{k+1}^{n^-}(x)} \Bigg]^2\\[2ex]
\intertext{leaving the asymptotic of $y_k(x)$ unchanged as $g_{1/2}(x)\to1$ for
$x\to\pm\infty$ with the shorthands}
  d_k^n (x) = \begin{cases}
    \ch(\pi x) + \cos(\pi \alpha^\prime) &\text{for } n+k \text{ even} \\
    \ch(\pi x) + \sin(\pi \alpha^\prime) &\text{for } n+k \text{ odd}
  \end{cases}\quad .
\end{gather}

\end{appendix}

\bibliographystyle{amsplain}
\bibliography{qtm,boundary}


\end{document}